# Non-destructive visualization of short circuits in lithium-ion batteries by magnetic field imaging system

Shogo Suzuki[1], Hideaki Okada[1], Kai Yabumoto[1], Seiju Matsuda[1], Yuki Mima[2], Noriaki Kimura[2], Kenjiro Kimura[1]

*Kobe University[1], 1-1, Rokkodai-cho, Nada-ku, Kobe-shi, Hyogo, 657-8501, Japan*

*Integral Geometry Science[2], 1-5-6, Minatojima-minamimachi, Chuo-ku, Kobe-shi, Hyogo 650-0047, Japan*

E-mail: staff-kimuralab@silver.kobe-u.ac.jp

To develop a high-density and long-life lithium-ion battery, a technology is needed that allows non-destructive visualization of the spatial distribution of deteriorated parts after cycle test. In the present study, we measured the distribution of the magnetic field leaking from the lithium-ion battery during its operation. Based on the measurement results, we evaluated the current density distribution inside a battery using the electric current reconstruction process. With respect to the changes in an internal state of the lithium-ion battery associated with cycle deterioration, we successfully visualized the spatial changes in the conductivity distribution inside the lithium-ion battery.

Because of their high energy density and long life, lithium-ion batteries are widely used in electric vehicles, hybrid electric vehicles, mobile phones, etc. Lithium-ion batteries, however, are also known for forming dendritic lithium crystals, which deposit on negative electrodes during charging[1-4]. Dendrites degrade the performance of the negative electrode and cause capacity deterioration[5-8]. In addition, it has been reported that some overgrown dendrites penetrate separators, which causes short-circuits that results in serious accidents such as ignition and burning of the organic solvent[9-12]. Observation of the negative electrode cross-section using synchrotron hard X-ray microtomography [13] and microscopic observation of specially shaped cells [14-16] demonstrated that the dendrites form due to the heterogeneous reactivity of the negative electrode. Accordingly, for a purpose of developing high-quality lithium-ion batteries, it is indispensable to establish a method for directly observing the phenomena occurring inside lithium-ion batteries to visualize the spatial non-uniformity of the reaction kinetics[17-20].

The methods for observing the non-uniformity of reactions include three-dimensional structural analysis using X-ray tomography[21, 22], visualization of the lithium ions distribution using X-ray Absorption Spectroscopy (XAS) [23-25], energy dispersive X-ray spectroscopy (EDXS) for element mapping[26, 27], and the Raman spectroscopy to visualize the crystal structure distribution of active materials in lithium-ion batteries[28, 29]. Also, in this study, in contrast to the mentioned methods, by utilizing the magnetic imaging system we measured the distribution of magnetic field generated by the currents during operation of a lithium-ion battery. Then, based on these results, we developed a method to visualize the conductivity distribution inside a lithium ion battery using the analytical relation of the inverse problem between the current in a battery and the magnetic field it induces. Therefore, this paper deals with a non-destructive visualization of changes in conductivity inside the lithium ion battery associated with its cycle deterioration.

Assigning the boundary conditions as the two-dimensional Fourier transform $f_x$ ($k_x$, $k_y$,) and $f_y$ ($k_x$, $k_y$,) of the measured magnetic field distribution in Eq. (1) and Eq. (2), the analytical solution of the basic equation of the static magnetic field in free space without a magnetic source can be derived by Eq. (3) and Eq. (4)[30]. As shown in Fig. 1, the *x*-axis and *y*-axis are in the electrode plane direction. With the magnetic field reconstruction method using this solution, the magnetic field distribution on the surface of the lithium ion battery can be obtained basing on the measurements.

$$f_x(k_x, k_y) = \int_{-\infty}^{\infty} \int_{-\infty}^{\infty} e^{-ik_x x - ik_y y} H_x(x, y, 0) dx dy \quad (1)$$

$$f_y(k_x, k_y) = \int_{-\infty}^{\infty} \int_{-\infty}^{\infty} e^{-ik_x x - ik_y y} H_y(x, y, 0) dx dy \quad (2)$$

$$H_x(x, y, z_0) =$$
$$= \frac{1}{(2\pi)^2} \int_{-\infty}^{\infty} \int_{-\infty}^{\infty} e^{ik_x x + ik_y y} f_x(k_x, k_y) e^{\sqrt{k_x^2 + k_y^2} z_0} dk_x dk_y \quad (3)$$

$$H_y(x, y, z_0) =$$
$$= \frac{1}{(2\pi)^2} \int_{-\infty}^{\infty} \int_{-\infty}^{\infty} e^{ik_x x + ik_y y} f_y(k_x, k_y) e^{\sqrt{k_x^2 + k_y^2} z_0} dk_x dk_y \quad (4)$$

Further, the relationship between the magnetic field distribution on the battery surface (Fig. 1) and conductivity inside the battery can be derived in the following manner[31]. In the equation (5), as shown in Fig. 1, $h_T$ is the distance between the electrodes, $h$ is the thickness of the electrode and $\sigma$ ($x$, $y$) is the conductivity distribution between the electrodes of a lithium ion battery, $\varphi(x, y)$ is the two-dimensional potential distribution on the electrode surface, $z_0$ is the electrode coordinate and $\sigma_0$ is the electrode conductivity.

$$\Delta H(x,y,z) = \begin{bmatrix} h_T^{-1} h \frac{\partial}{\partial y}(\sigma(x,y)\varphi(x,y))\delta(z-z_0) - \sigma_0 h \frac{\partial \varphi(x,y)}{\partial y} \delta'(z-z_0) \\ h_T^{-1} h \frac{\partial}{\partial x}(\sigma(x,y)\varphi(x,y))\delta(z-z_0) - \sigma_0 h \frac{\partial \varphi(x,y)}{\partial x} \delta'(z-z_0) \\ 0 \end{bmatrix} \quad (5)$$

$Q_x(k_x, k_y, z_0)$ and $Q_y(k_x, k_y, z_0)$ are two-dimensional Fourier transforms of the $x$ and $y$ components of the magnetic field on the battery surface according to Eq. (6) and Eq. (7).

$$Q_x(k_x, k_y, z_0) = \int_{-\infty}^{\infty} \int_{-\infty}^{\infty} e^{-ik_x x - ik_y y} H_x(x, y, z_0) dx dy \quad (6)$$

$$Q_y(k_x, k_y, z_0) = \int_{-\infty}^{\infty} \int_{-\infty}^{\infty} e^{-ik_x x - ik_y y} H_y(x, y, z_0) dx dy \quad (7)$$

Using Eq. (6) and Eq. (7), $\varphi(x, y)$ can be derived from Eq. (5) as shown in Eq. (8)[31]. Furthermore, the analytical solution $\sigma(x, y)$ of the two-dimensional conductivity distribution in the battery can be obtained from Eq. (9). This distribution represents the electric current between the electrodes, and corresponds to the reaction rate of the active material and the ion diffusion rate in the electrolyte solution. From the magnetic field distribution on the surface determined by this method, the conductivity distribution inside the battery can be reconstructed.

$$\varphi(x,y) = \frac{1}{(2\pi)^2} \int_{-\infty}^{\infty} \int_{-\infty}^{\infty} e^{ik_x x + ik_y y} \frac{2\{ik_y Q_x(k_x, k_y, z_0) - ik_x Q_y(k_x, k_y, z_0)\}}{h\sigma_0(k_x^2 + k_y^2)\left(h\sqrt{k_x^2 + k_y^2} - 1\right)} dk_x dk_y \quad (8)$$

$$\sigma(x,y) = hh_T \sigma_0 \frac{\left(\frac{\partial^2}{\partial x^2} + \frac{\partial^2}{\partial y^2}\right)\varphi(x,y)}{\varphi(x,y)} \quad (9)$$

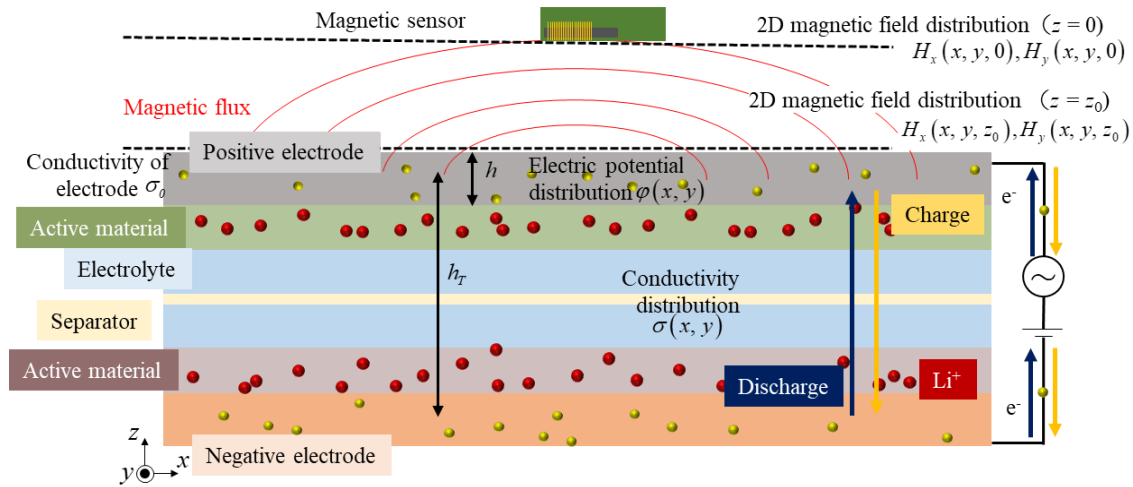

**Fig. 1**. Schematic diagram for measuring the magnetic field during the operation of a lithium-ion battery.

Fig. 2 shows a schematic diagram of the magnetic imaging system. We used a magnetic sensor (Detectability : 30 pT /Hz$^{0.5}$ at 1 Hz) that was developed based on the magneto-impedance effect[32]). The magnetic field distribution was measured by controlling the X and Y stepping motors and performing two-dimensional scanning. Further, the angle of the magnetic sensor was adjusted by the θ stepping motor to measure the *x* component $H_x$ and the *y* component $H_y$ of the magnetic field. When applying the alternating current with the 1 Hz frequency source to the battery, the generated magnetic field was detected by the magnetic sensor. After digitalization of the magnetic sensor signals by a 16-bit A/D converter, magnetic field was phase detected with the reference signal of the current source. Then, the conductivity distribution in the lithium-ion battery was reconstructed following Eq. (4) and Eq. (9).

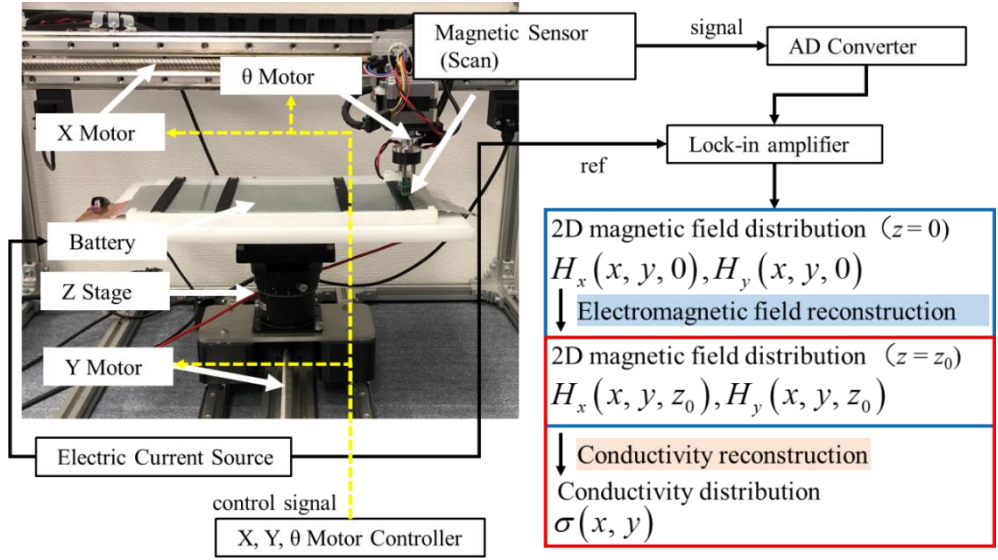

**Fig. 2.** Magnetic imaging system and overview of the process to visualize the conductivity distribution in a lithium-ion battery.

In the cycle test, we used a laminated single-layer lithium-ion battery having an electrode size of 80 mm × 240 mm. The cathode consisted of NMC622 (NMC: Nickel-Manganese-Cobalt cathode) active material, acetylene black (AB) as a conductive additive, and polyvinylidene fluoride (PVDF) at the ratio of 94:3:3. The anode consisted of graphite as active material, AB, carboxymethylcellulose (CMC) as a thickener, and styrene-butadiene rubber (SBR) as a binder at the ratio of 98.5:0.5:1:1. The electrolyte was 1.0 M lithium hexafluorophosphate (LiPF$_6$) solution in ethylene carbonate (EC) / dimethyl carbonate (DMC) / ethyl methyl carbonate (EMC) at the ratio of 1/1/1, with 1% doping with vinylene carbonate (VC). The cycle test was performed at 15 °C. The battery was charged and discharged by DC. The charging current was 2250 mA (5 C) , the control voltage was 4.2 V, and the stop condition was set at 45 mA (0.1 C). The discharging current was 450 mA (1 C) and the stop voltage was set at 2.5 V. The capacity of the battery before the cycle test was 469 mAh, and those after 100 and 200 cycles were 419 mAh and 252 mAh, respectively, with the progressive decrease in capacity upon repeated cycle tests. The capacity was measured at the test temperature of 25 °C. The battery was charged by constant-current/constant-voltage charging, and discharged by constant-current discharging. The charging/discharging conditions were 45 mA (0.1 C) for constant current charging, 4.2 V for constant voltage charging control voltage, 9 mA (0.02 C) for stop condition, 45 mA (0.1 C) for constant current discharge and 2.5 V for stop voltage. The magnetic field distribution was measured before the cycle test, after 100 cycles, and after 200 cycles. The magnetic field was measured by applying to the battery the AC (1 Hz, 240 mAp-p) superimposed with the DC voltage of 3.4 V. Using the non-destructive magnetic imaging system illustrated in Fig. 2, measuring magnetic field was performed over the area of 260 mm × 120 mm divided into 32 × 16 pixels, and measurement time of 25 sec each. By processing the difference between the magnetic field distributions before and after the cycles following Eq. (9), we visualized the changes in the conductivity distribution during the cycle test. This procedure makes it possible to visualize only the changes in conductivity inside the battery associated with cycle deterioration, and to neglect the deviation of initial current density distribution. Figs. 3 (a) and 3(b) show the results of spatial distribution of $H_x$ and $H_y$ components of the magnetic field leaking from the lithium-ion battery before the cycle test. Figs. 3 (c) and 3(d) show the spatial distribution of $H_x$ and $H_y$ after the cycle test (100 times), Figs. 3(e) and 3(f) show the difference between the spatial distributions of $H_x$ and $H_y$ on the battery surface before and after the 100 cycles. Fig. 3 (g) visualizes the conductivity distribution basing on the data in Figs. 3(e) and 3(f) calculated using Eq. (9). Furthermore, Figs. 3(h) and 3(i) show the spatial distribution of $H_x$ and $H_y$ before and after the 200 cycles, and Figs. 3(j) and 3(k) show the difference between the spatial distributions $H_x$ and $H_y$ on the battery surface before and after the 200 cycles. Moreover, Fig. 3(l) visualizes the conductivity distribution basing on the data in Figs. 3(j) and 3(k). Following Figs. 3(g) and 3(l), it is clear that the changes associated with the cycle test were most significant in the center of the battery, and the region of abnormal conductivity expands during the cycle tests.

In this study, we visualized the spatial changes in the conductivity distribution inside a Li-ion battery due to cycle deterioration by measuring the magnetic field generated by the current flowing during its operation. In addition, we successfully identified the deteriorated part causing the battery capacity decrease during the charging-discharging cycles. Furthermore, as this system provides a non-destructive analysis, the defective part size and shape can be visualized during the cycle tests. Thus, it was demonstrated that the developed system was efficient for visualizing the short-circuits occurred in lithium-ion batteries.

## Acknowledgments

This work was partially supported by New Energy and Industrial Technology Development

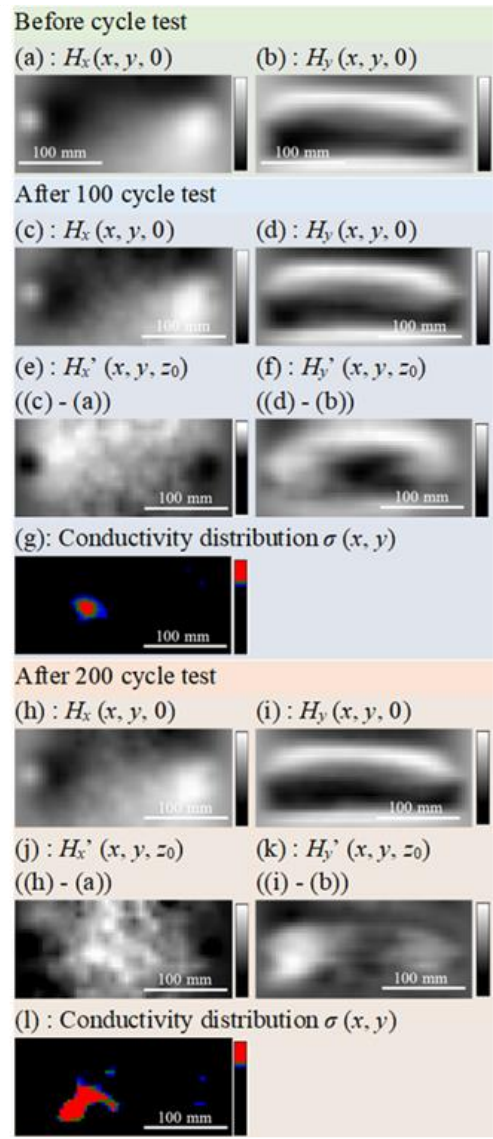

**Figs. 3** (a) magnetic field distribution before cycle test ($H_x$ (x, y, 0)) obtained by magnetic imaging system; (b) magnetic field distribution before cycle test ($H_y$ (x, y, 0)); (c) magnetic field distribution after 100 cycles ($H_x$ (x, y, 0)); (d) magnetic field distribution after 100 cycles ($H_y$ (x, y, 0)); (e) difference in magnetic field distribution ($H_x$' (x, y, $z_0$)) before and after 100 cycles by applying the magnetic field reconstruction method; (f) difference in magnetic field distribution ($H_y$' (x, y, $z_0$)) before and after 100 cycles by applying the magnetic field reconstruction method;

Organization (NEDO).

(g) conductivity distribution ($\sigma(x, y)$) obtained from the data of the measurements in (e) and (f) using Eq. (9); (h) magnetic field distribution after 200 cycles ($H_x(x, y, 0)$); (i) magnetic field distribution after 200 cycles (200 times) ($H_y(x, y, 0)$); (j) difference in magnetic field distribution ($H_x'(x, y, z_0)$) before and after 200 cycles by applying the magnetic field reconstruction method; (k) difference in magnetic field distribution ($H_y'(x, y, z_0)$) before and after 200 cycles by applying the magnetic field reconstruction method; (l) conductivity distribution ($\sigma(x, y)$) obtained from the data of the measurements in (j) and (k) using Eq. (9).